\documentclass[prd,nofootinbib,twocolumn,superscriptaddress,preprintnumbers,balancelastpage,nofootinbib]{revtex4-1}

\usepackage{graphicx}  
\usepackage{dcolumn}   
\usepackage{bm}        
\usepackage{amssymb, amsmath}
\usepackage{slashed}   

\usepackage[usenames,dvipsnames,svgnames,table]{xcolor}

\def\be{\begin{equation}}
\def\ee{\end{equation}}
\def\bea{\begin{eqnarray}}
\def\eea{\end{eqnarray}}

\newcommand{\mone}{m_{12}}
\newcommand{\mtwo}{m_{23}}
\newcommand{\GeV}{{\rm GeV}}

\def\beq{\begin{equation}}
\def\eeq{\end{equation}}

\def\be{\begin{eqnarray}}
\def\ee{\end{eqnarray}}

\begin{document}
\widetext

\newcommand\Princeton{ Princeton University, Princeton, NJ }
\newcommand\FNAL{Fermi National Accelerator Laboratory, Batavia, IL }

\preprint{FERMILAB-PUB-18-665-A}
\preprint{KEK-TH-2105}

\title{Probing Muon-Philic Force Carriers and Dark Matter at Kaon Factories}
\author{Gordan Krnjaic}
\affiliation{\FNAL}
\author{Gustavo Marques-Tavares}
\affiliation{Maryland Center for Fundamental Physics, Department of Physics, \\
	University of Maryland, College Park, MD 20742}
\affiliation{Stanford Institute for Theoretical Physics, \\
	Stanford University, Stanford, CA 94305, USA}
\author{Diego Redigolo}
\affiliation{Tel-Aviv University, Tel-Aviv Israel}
\affiliation{Institute for Advanced Study, 
Princeton, NJ USA}
\affiliation{ Weizmann Institute of Science, Rehovot Israel}
\author{Kohsaku Tobioka}
\affiliation{Florida State University, Tallahassee, FL USA}
\affiliation{High Energy Accelerator Research Organization (KEK), Tsukuba  Japan}

\begin{abstract}
\noindent
 
Rare kaon decays are excellent probes 
of light, new weakly-coupled particles. If such particles $X$ couple preferentially to muons, 
they can be produced in $K\to \mu \nu X$ decays. In this Letter we evaluate the future sensitivity for this process at NA62 assuming $X$ decays either invisibly or to di-muons. 
Our main physics target is the parameter space that resolves the $(g-2)_\mu$ anomaly, where $X$  is a gauged $L_\mu-L_\tau$ vector or a muon-philic scalar. 
The same parameter space can also accommodate dark matter freeze out or reduce the tension between cosmological and local measurements of $H_0$ if the new force decays to dark matter or neutrinos, respectively. We show that for invisible $X$ decays, a dedicated single muon trigger analysis at NA62 could probe much of the remaining $(g-2)_\mu$ favored parameter space.  
Alternatively, if $X$ decays to muons, NA62 can perform a di-muon resonance search in $K\to 3 \mu \nu$ events and greatly improve existing coverage for this process. 
Independently of its sensitivity to new particles, we find that NA62 is also sensitive to the Standard Model predicted rate for $K \to 3\mu \nu$, which has never been measured.
\end{abstract}

\maketitle
\noindent




\section{Introduction}
Light weakly-coupled forces arise in many compelling extensions of the Standard Model (SM) 
and are the focus of a broad experimental effort \cite{Essig:2013lka,Battaglieri:2017aum,Alexander:2016aln}. 
 If the corresponding force-carriers \footnote{We refer to force carriers as ``forces" throughout.} couple preferentially to muons, they offer the last viable opportunity to 
 resolve the longstanding $\sim 3.5 \, \sigma$ anomaly in $(g-2)_\mu$ \cite{Bennett:2006fi, Hagiwara:2011af, Davier:2010nc} with new physics below the electroweak scale as proposed in \cite{Pospelov:2008zw}.\footnote{Light new particles with appreciable
 couplings to the first generation have been excluded in simple models, including both visibly and invisibly decaying dark photons (see \cite{Alexander:2016aln,Mohlabeng:2019vrz}).} Thus, there is strong motivation to improve experimental sensitivity to these interactions. 
 
 Furthermore, there is abundant evidence for the existence of dark matter (DM), 
 whose microscopic properties remain elusive~\cite{Bertone:2016nfn}.  
 One possible explanation for these null results is that DM couples more strongly to the second
 and third generation. Indeed, there are several consistent, viable,
 and {\it predictive} dark forces which mediate DM freeze-out to 
  higher generation particles~\cite{Agrawal:2014ufa,Kahn:2018cqs}. Since muonic forces don't couple directly to first generation particles, these DM candidates
are difficult to probe with direct detection experiments, but can be efficiently produced at accelerators.  

\begin{figure*}[t!]
\hspace{-0.5cm}
\includegraphics[width=0.46\textwidth]{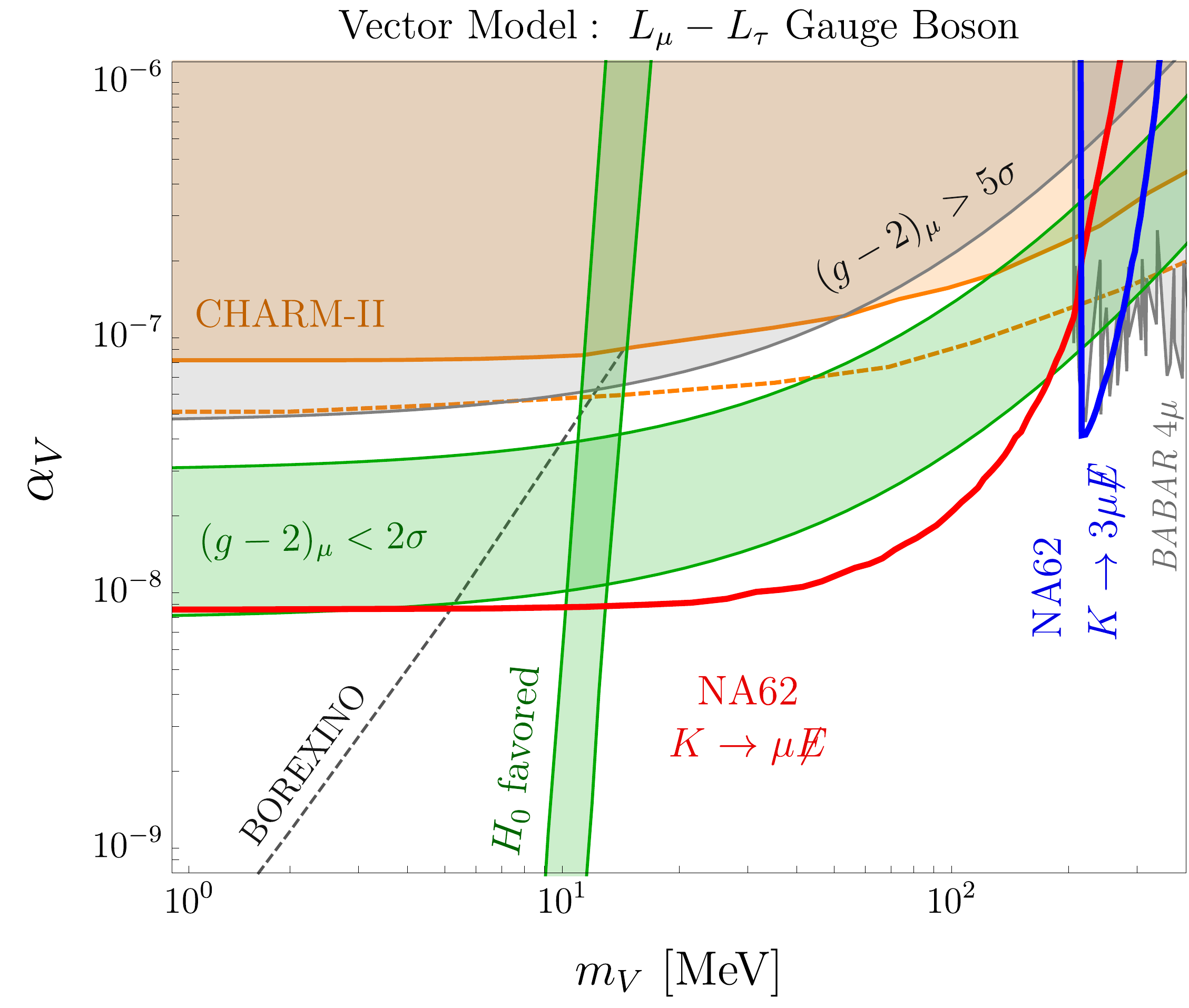}~~~
\includegraphics[width=0.46\textwidth]{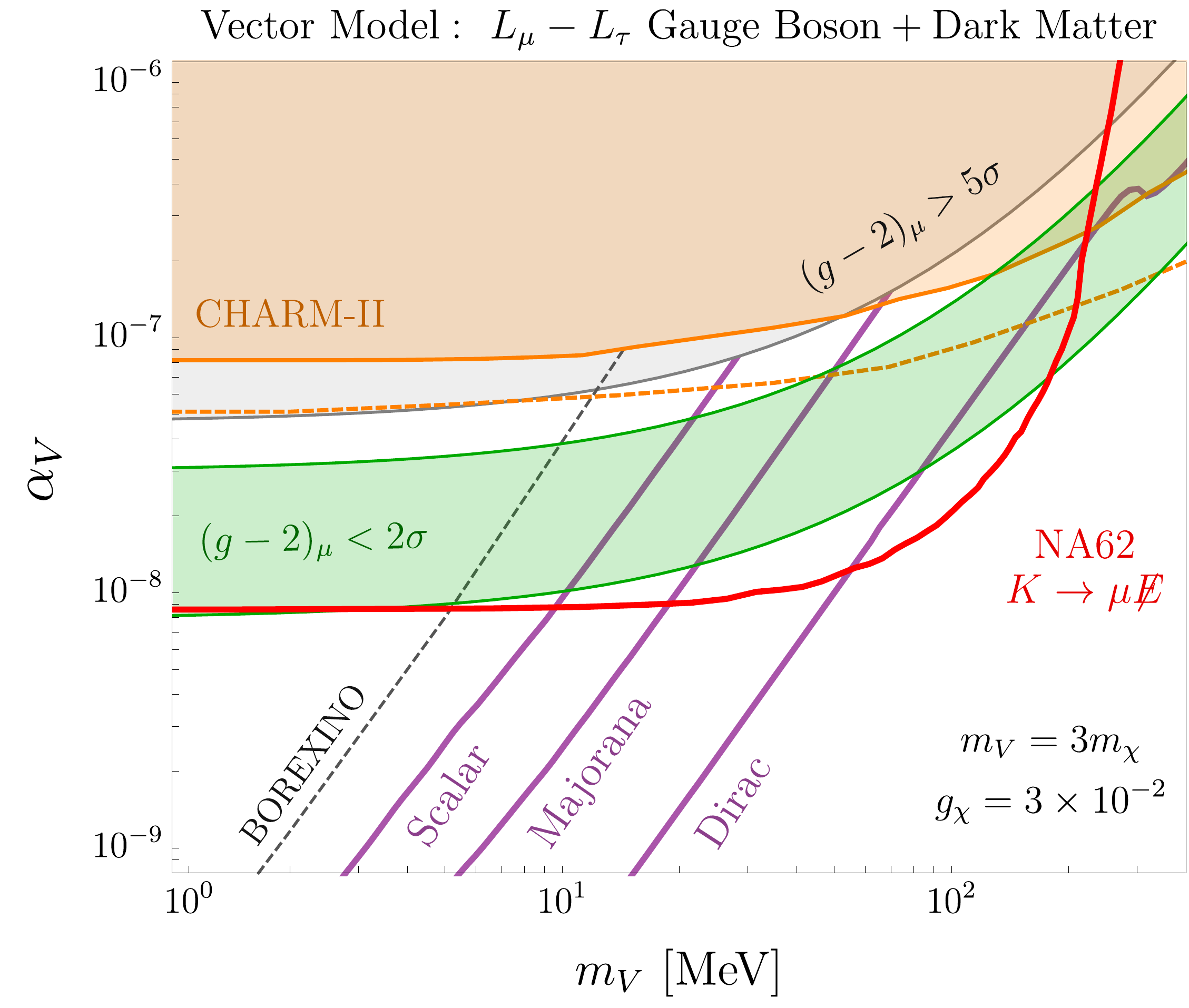}
\caption{
{\bf Left:}
Parameter space for an $L_{\mu}-L_{\tau}$ SM extension from Sec.~\ref{sec:vector}. The light green band is the $2\sigma$ region accommodating the $(g-2)_\mu$ anomaly, while the green vertical region increases $\Delta N_{\rm eff} = 0.2-0.5$, ameliorating the $H_0$ tension \cite{Escudero:2019gzq}. We show  projections for an NA62 search for $K^+ \to \mu^+ \nu_\mu V$ followed by a prompt invisible $V \to \nu  \bar\nu$ decay (red curve) or a prompt visible $V \to \mu^+\mu^-$ decay (blue curve). Both sensitivities assume the full NA62 luminosity to be recorded by the single muon and di-muon trigger respectively and systematic errors comparable to the statistical uncertainty (see Sec.~\ref{sec:analysis} and the supplementary material~\cite{supplement} which contains Refs. 
\cite{Carlson:2012pc,CortinaGil:2018fkc,CortinaGil:2017mqf,Chen:2015vqy,Batell:2016ove,Batell:2017kty,Chen:2017awl,Krnjaic:2015mbs}
 for details). We also show bounds from {\it BABAR} $4\mu$, \cite{TheBABAR:2016rlg}, $(g-2)_\mu$, and CHARM-II $\nu$ \cite{Geiregat:1990gz,Altmannshofer:2014pba}; the dashed curve is the CCFR bound \cite{Mishra:1991bv}. The dashed Borexino bound \cite{Bellini:2011rx,Harnik:2012ni,Kaneta:2016uyt} assumes a the mixing from SM loops.
{\bf Right:} Same as left, only the $V$ decays to dark matter $\chi$, with ${\rm BR}(V \to \chi \chi) \simeq 1$; the purple bands  yield the observed DM abundance  via freeze out.}
\label{fig:money}
\end{figure*}

It is known that muonic forces lead to new rare kaon decays \cite{Reece:2009un,Batell:2016ove,Ibe:2016dir}. However, there are several timely reasons to revisit this subject:
\begin{enumerate}
\item  The NA62 experiment \cite{Martellotti:2015kna} is currently producing unprecedented numbers of kaons, and is poised to considerably improve sensitivity to muonic forces. 
\item  The $g-2$ collaboration \cite{Grange:2015fou} and the J-PARC  $g-2$ experiment \cite{Mibe_2010} will soon decisively test the $(g-2)_\mu$ anomaly. If this discrepancy is due to new physics, the particles responsible necessarily predict SM deviations in other, complementary muonic systems.
\item  Recently there has been great interest in new proposals for dedicated experiments to probe muonic forces \cite{Abbon:2007pq,Gninenko:2014pea,Chen:2017awl,Kaneta:2016uyt,Kahn:2018cqs}.
To  assess the merits of these ideas, it is essential to know what existing experiments can achieve.
\end{enumerate}
In this Letter we show that existing kaon factories, can probe $K\to \mu \nu X$ decays where $X$ is a new particle that couples preferentially to muons. Our main focus are the new physics opportunities of the NA62 experiment at CERN \cite{NA62status}, which will produce ~$10^{13}$ $K^+$.  

If $X$ decays invisibly, we find that, with a dedicated single muon trigger, NA62 could have unprecedented sensitivity to $K\to \mu\nu X (X \to$ invisible) processes. Such a search could probe 
nearly all the remaining parameter space in which muonic forces reconcile the $(g-2)_\mu$ anomaly. If the invisible particles are DM, this
also enables $X$-mediated thermal freeze out \cite{Kahn:2018cqs}; if, instead, these particles are neutrinos, this
same parameter space can ease the $\sim 3.5 \sigma$ tension in Hubble constant measurements \cite{Escudero:2019gzq}.
 
If $X$ decays to muons, we find that an NA62 di-muon resonance search in $K\to \mu\nu X( X\to \mu^+\mu^-)$ processes could greatly improve the coverage for both scalar and vector forces, thereby covering nearly all of the $(g-2)_\mu$ favored region for $m_K - m_\mu > m_X > 2m_\mu$. The irreducible background for this search arises from $K\to 3 \mu \nu$ decays which have never been observed before; intriguingly, we find that NA62 can also measure this process in existing data.




\section{Vector Forces}
\label{sec:vector}

\subsection{Gauged $L_\mu - L_\tau$}
A vector $V$ gauging a spontaneously broken $L_\mu - L_\tau$ symmetry is a minimal candidate to explain the $(g-2)_\mu$ anomaly.
 The Lagrangian contains   
\be
\label{eq:lagrangian}
{\cal L} \supset  \frac{m_{V}^2}{2} V_\mu V^\mu + V_\mu \left(  g_V  J_V^{\mu} + \epsilon e J_{\rm \tiny EM}^\mu \right),~~~
\ee
where $g_V$ is the gauge coupling, $m_{V}$ is the mass, and $J_V^{\mu}$ is the  $L_\mu -L_\tau$ current~\cite{He:1991qd}. Loops of taus and muons induce kinetic mixing with the  photon $\epsilon\simeq g_V/67$,
which also couples $V$ to the EM current $J_{\rm EM}^\mu$ in Eq.~(\ref{eq:lagrangian}). 
 The widths for $V\to f\bar f$ are
\begin{align}
&\Gamma_{V \to f \bar{f}}  =  \frac{\alpha_V m_{V}}{3}\left(1+\frac{2m_\mu^2}{m_V^2}\right)\sqrt{1-\frac{4 m_\mu^2}{m_V^2}}\ , 
\end{align}
where $f = \mu,\tau$ and $\alpha_V \equiv g_V^2/4\pi$, and the width to neutrino flavor $\nu_f$ is 
$\Gamma_{V \to \nu_f \bar \nu_f}  =  \alpha_V m_{V}/6$.  Decays through the EM current are suppressed by additional factors of $\epsilon^2 \alpha/\alpha_V$, so we neglect these here. In all of the parameter space we consider here,  $V$ decays promptly within the 65~m decay region of NA62.  


%
  
Although we require $m_V \gtrsim 1 $ MeV to avoid tension with cosmology~\cite{Pospelov:2010hj}, for $m_V \sim $ few MeV, $V \to \nu \bar \nu$ decays after neutrino decoupling increase the effective number of neutrino species by $\Delta N_{\rm eff} \sim 0.2-0.5$, which can ameliorate the tension in Hubble rate measurements \cite{Escudero:2019gzq}; lighter masses are disfavored \cite{Kamada:2015era,Aghanim:2018eyx}.

As shown in Fig.~\ref{fig:money} (left), the NA62 $K\to \mu\nu X$ reach with $X$ decaying invisibly could cover a large portion of the parameter space, far beyond the reach of present experiments. Conversely the $K\to \mu\nu X$ search with $X\to \mu\mu$ is competitive with {\it BABAR}. The detailed study and the experimental challenges of the invisible and di-muon analyses are described in Sec.~\ref{invisible} and Sec.~\ref{sec:dimuon} respectively. 

\subsection{Adding $L_\mu - L_\tau$ Charged Dark Matter}

If DM couples to $V$, $V\to$ DM decays can significantly change the $V$ branching fraction above the di-muon threshold; below this boundary,  $V$ always decays invisibly (either to neutrinos or DM). Here we add a  DM candidate $\chi$ $(m_V > 2m_\chi)$ charged under $L_\mu- L_\tau$ and 
extend Eq.~(\ref{eq:lagrangian})  to include a coupling to the dark current
${\cal L } \supset g_{\chi} V_\mu J_\chi^\mu$.
We now have
\be
\label{eq:dmcurrents}
J_\chi^\mu= 
\begin{cases}
                        i \chi^* \partial_\mu  \chi + h.c. ~ & {\rm Complex~ Scalar} \\
              \frac{1}{2}  \overline \chi \gamma^\mu \gamma^5 \chi ~  &{\rm Majorana} \\
                     \overline \chi \gamma^\mu \chi ~ & {\rm Dirac} 
                                         \end{cases}
\ee
where $g_\chi \equiv g_V q_\chi$ is the DM-$V$ coupling and  $q_\chi$ is the DM $L_\mu-L_\tau$ charger; we assume $\mu,\tau$ and $\nu_{\mu,\tau}$ carry unit charge. 
For $m_\chi < m_V$, freeze out proceeds via $s$-channel annihilation to SM particles for each model in Eq.~(\ref{eq:dmcurrents})~\cite{Kahn:2018cqs,Berlin:2018sjs}. Figure~\ref{fig:money} shows DM production targets alongside various constraints.
 \begin{figure*}[t!]
\hspace{-0.5cm}
\includegraphics[width=0.46\textwidth]{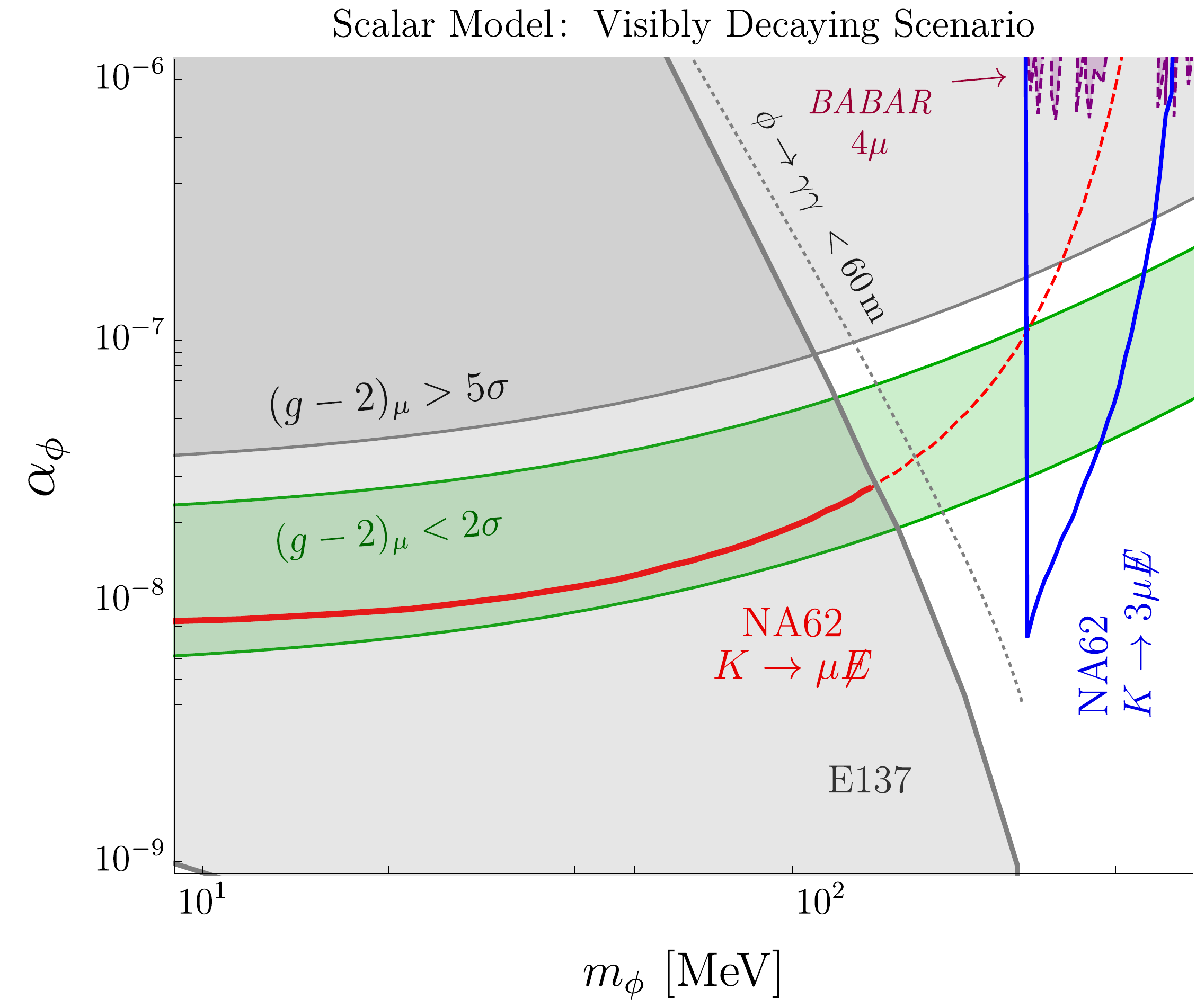}~~
\includegraphics[width=0.46\textwidth]{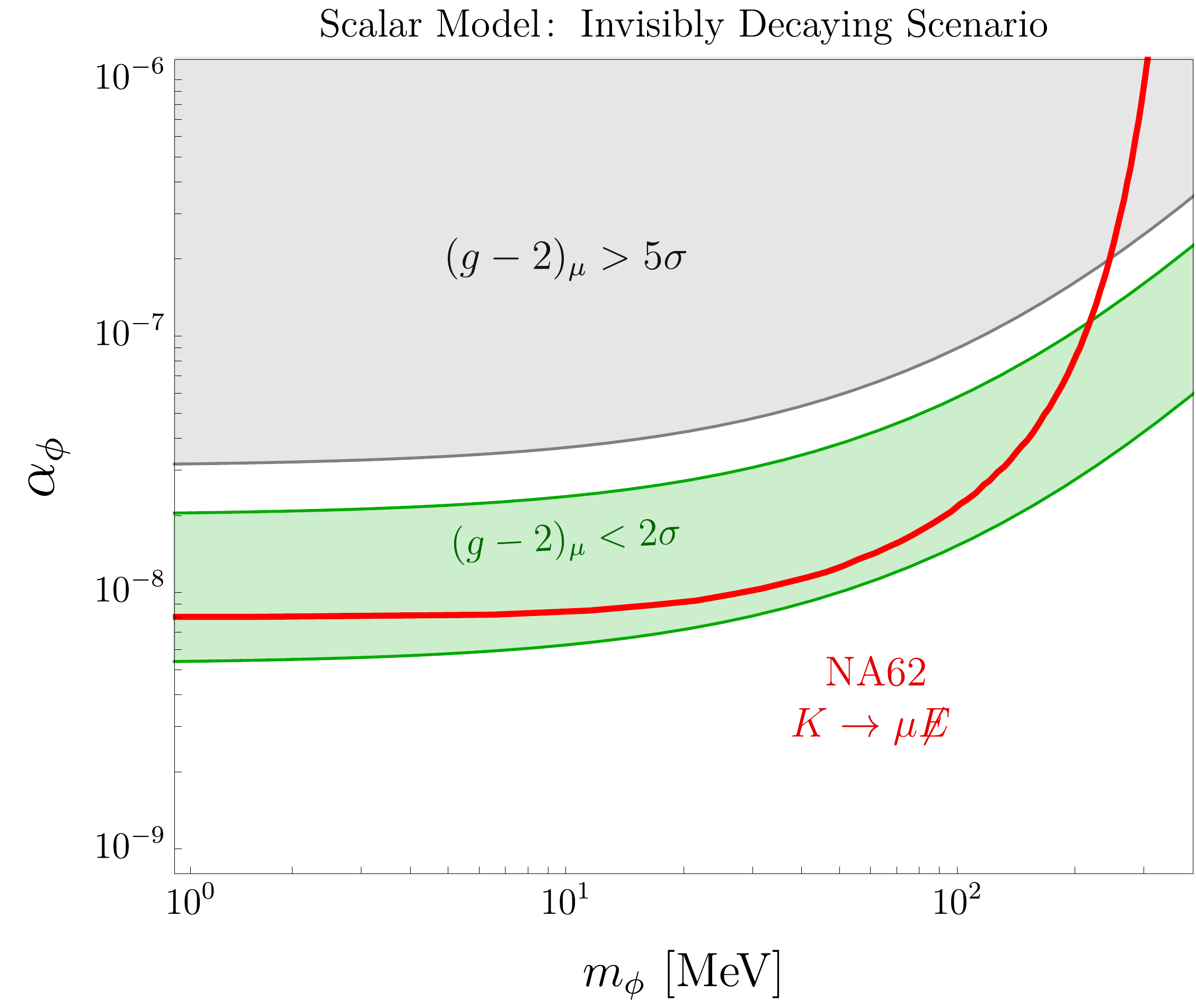}~~
\caption{
Parameter space and NA62 projection for a muon-philic scalar particle $\phi$ described in Sec.~\ref{sec:scalar}.
Here we define $\alpha_\phi \equiv y_\phi^2/4\pi$ where $y_\phi$ is the Yukawa coupling to muons
from Eq.~(\ref{eq:yukawa}) and the light green band accommodates the $(g-2)_\mu$ anomaly. {\bf Left:} Projections for an NA62 search for $K^+ \to \mu^+ \nu_\mu \phi$ where $\phi$  decays  visibly into $\phi \to \mu^+\mu^-$ or $\gamma \gamma$ where kinematically allowed. On the left of the dashed grey line the lifetime of the muon-philic scalar is long enough to give an invisible signal at NA62. 
Also shown are E137 constraints from \cite{Marsicano:2018vin}.
{\bf Right:} Same as the left, but $\phi$ decays invisibly. 
Both assume the full NA62 luminosity and the searches to be statistics dominated (see Sec.~\ref{sec:analysis} and supplementary material~\cite{supplement} for details). 
}
\label{fig:scalar}
\end{figure*}




\section{Scalar Forces}
\label{sec:scalar}
The minimal Lagrangian for a Yukawa muonic force is 
\be
\label{eq:yukawa}
{\cal L} = \frac{1}{2} (\partial_\mu \phi)^2 - \frac{m_\phi^2}{2} \phi^2 - y_\phi \phi \bar \mu \mu,
\ee
where $\phi$ is a real scalar particle. The interaction in Eq.~(\ref{eq:yukawa})  
can arise, for instance, by integrating out a heavy, vectorlike lepton singlets whose mass mixes with 
the right handed muon as discussed in the supplementary material~\cite{supplement}. In the absence of additional interactions, for $m_\phi > 2m_\mu$, 
the dominant decay is $\phi \to \mu^+\mu^-$ with partial width
\be
\Gamma_{\phi \to \mu^+\mu^-} = \frac{\alpha_\phi m_\phi}{2} \left( 1 - \frac{4m_\mu^2}{m_\phi^2}\right)^{3/2},
\ee
where $\alpha_\phi \equiv y_\phi^2/4\pi$. For $m_\phi < 2m_\mu$, the dominant channel is $\phi \to \gamma \gamma$ through a muon loop
with width 
\begin{equation}
\Gamma_{\phi \to \gamma\gamma} =  \!\frac{\alpha^2_{\text{EM}}  \alpha_\phi  m_\phi^3 }{ 64 \pi^2  m_\mu^2}    \left| \frac{2}{x^2} \bigl( x + (x-1)   \arcsin^2 \!\!  \sqrt{x}   \,\, \bigr)  \right|^{2}\!\!,~~~~
\label{eq:width-to-diphoton}
\end{equation}
where $x \equiv m_\phi^2/4m_\mu^2$ and the lab frame decay length is 
\be
\!\!\ell_{\phi \to \gamma\gamma} \sim  60 {\rm m} \left(\frac{3 \times 10^{-6}}{ \alpha_\phi}\right)\left(\frac{\rm 50\, MeV}{m_\phi}\right)^4 \left(\frac{E_\phi}{\rm 75\, GeV}\right)\!\!,~~
\ee
where the $m_\phi^{-4}$ scaling accounts for the boost factor.
 In this minimal ``visibly decaying" scenario,
 most of our favored parameter space is below the di-muon threshold, so the diphoton channel dominates and, for the maximum $\phi$ energy $\sim 75$ GeV, 
nearly all  decays occur outside the NA62 detector to mimick a missing energy signature. However, a dedicated study is required to
identify the distance beyond which these decays are invisible given NA62 kinematics and acceptance; we also note that it may be possible to
perform a $\phi\to\gamma\gamma$  resonance search if this occurs inside the decay region.
 
 Alternatively, $\phi$ may decay predominantly to undetected particles (e.g DM) in the ``invisibly decaying" scenario. In both cases, the scalar is produced via $K\to \mu \nu \phi$ processes whose width is computed in the supplementary material~\cite{supplement}.  
 
Figure~\ref{fig:scalar} shows the NA62 projections for visible (left) and invisible (right) decays assuming $100\%$ branching ratio in both channels. The main difference relative to the vector case is that the $K\to 3\mu\nu$ search improves considerably beyond the {\it BABAR} $4\mu$ bounds; here the $e^{+}e^{-}\to \mu^{+}\mu^{-}\phi$ cross section is much smaller 
 than $V$ production.  We also show the E137 bound for visible decays from~\cite{Marsicano:2018vin}~(see also \cite{Dolan:2017osp}). There are additional constraints from supernovae~\cite{Chen:2018vkr,Marsicano:2018vin} not included in the figure due to their large astrophysical uncertainties and significant model dependence in the invisible decaying scenario.




\section{Rare kaon Decays at NA62}\label{sec:analysis}

The electroweak coupling governing SM $K\to \mu  \nu$ decays  is
\be
	\mathcal L &\supset& (2 G_F f_K \, V_{us}) \, \partial_\alpha K^- \bar \nu_\mu \gamma^\alpha P_L \mu  + h.c., ~~~~~
\label{effectivInt-K}	
\ee
where 
	  $G_F$ is the Fermi constant, $V_{us} = 0.223$ is the $us$ CKM  element, and $f_K = 160$ MeV is the kaon decay constant.
	We are interested in three-body corrections to this process: $K^+ \to \mu^+ \nu_\mu X$, where  $X  = V$ or  $\phi$, is 
emitted from a final state $\mu$ and/or $\nu_\mu$ line. The differential decay distribution is
\be
\label{eq:differential}
	\frac{d \Gamma(K^+\to \mu^+\nu X)}{  dm_{\rm \tiny miss}^2 \!\! \! } \! =  \frac{1}{256 \pi^3 m_K^3} \int \sum |  \mathcal{M} |^2 dm_{\mu X}^2,~~~
\ee
where $m_{\mu X}$ is the $\mu X$ invariant mass and 
\begin{equation}
m^2_{\text{miss}}\equiv (P_X+P_{\nu_\mu})^2 =(P_K-P_\mu)^2\, .
\end{equation}
The matrix element $|{\cal M}|^2$  for both scenarios  is calculated in the supplementary material~\cite{supplement}. Below we describe two different search strategies depending on 
whether X decays invisibly or to muons.  

\subsection{Invisible analysis}\label{invisible} 
If $X$ is produced in $K^+ \to \mu^+ \nu_\mu X$ events and decays {\it invisibly}, the $m^2_{\text{miss}}$ distribution in
$K \to \mu + $ invisible decays differs from the SM prediction (see supplement~\cite{supplement}). The  sensitivity of an $m^2_{\text{miss}}$ search in single muon events is computed using the log-likelihood ratio 
\begin{align}
\Lambda(S)=\sum_{i} -2\log \frac{L_i(S)}{L_i(\hat{S}=0)}\ ,    \label{loglike}
\end{align}
where $L_i$, the likelihood in each bin $i$, is constructed from a Poisson distribution,\footnote{$L_i(S)=\frac{(S\epsilon_{Si}+B_i)^{D_i}}{D_i!}e^{-(S\epsilon_{Si}+B_i)}$ where $D_i$, $B_i$, and $\epsilon_{Si}$ are data, background, and signal fraction in each bin. The maximum likelihood estimator is $\hat S=0$ under the assumptions behind our projections, $D_i=B_i$.} and $S= N_{K^+} \, \mathcal{A} \,  \text{BR}(K^+\to \mu^+\nu X)$ is the signal yield with acceptance $\mathcal{A}\simeq0.35$. 
We require $\Lambda(S) < 4$ to define the $2\sigma$ sensitivity. 

Our background sample is extracted from public NA62 data from the 2015 run in which $2.4 \times 10^7$ events passing the single muon trigger were recorded~\cite{CortinaGil:2017mqf}. These data yield $N_{K^+}\approx 10^{8}$ kaons after dividing out the detector acceptance and SM branching ratio $\text{BR}(K^+ \to \mu^+ \nu_\mu)=0.63$; all events in this sample are binned in missing mass intervals of $ 4 \times 10^{-3}\text{ GeV}^2$.  

One of the main backgrounds for this search is $K\to\mu\nu(\gamma)$, in which a radiated $\gamma$ is not detected and contributes to the missing energy. This process peaks at $m^2_{\rm miss} = 0$ and its contribution to  the large missing mass tail depends on NA62's photon rejection efficiency. Due to this large background, including missing mass bins below $m_{\text{miss}}^2=2.3\times 10^{-2}\text{ GeV}^2$ does not change the log-likelihood ratio defined in Eq.~\eqref{loglike}.

In the 2015 data sample, other backgrounds are present at large $m_{\rm miss}^2$ and exceed the $K\to\mu\nu(\gamma)$ tail for $m^2_{\text{miss}}>0.1\text{ GeV}^2$. These events are largely due to the muon halo
and we expect their contribution to be substantially reduced in the 2017 dataset where NA62 utilizes a silicon pixel detector (GTK) to measure the timing and momentum of upstream Kaons \cite{NA62status}. 
To approximately account for this existing improvement, we rescale the background yield above $m_{\text{miss}}^2>2.3\times 10^{-2}\text{ GeV}^2$ by an additional factor of two to estimate our sensitivity.\footnote{We thank E. Goudzovski and B. D$\ddot{\text{o}}$brich for discussions on this.}   For more details regarding our analysis and the challenges of maximizing
signal sensitivity, see supplementary material~\cite{supplement} where we show how our results vary under different assumptions regarding systematic errors.

\subsection{Di-muon analysis}\label{sec:dimuon}

If $X$ is produced in $K^+ \to \mu^+ \nu_\mu X$ events and decays {\it visibly} to di-muons, NA62 can improve upon previous experiments in the $K^+ \rightarrow 2 \mu^+  \mu^- \nu$ channel. The SM prediction is $ {\rm BR}(K^+ \rightarrow 2\mu^+ \mu^-  \nu )_{\rm SM} = 1.3 \times 10^{-8}$~\cite{Bijnens:1992en} and currently has not been observed;  the best limit comes from E787  ${\rm BR}(K^+ \rightarrow 2\mu^+\mu^- \nu  )_{\rm obs} < 4.1 \times 10^{-7}$ in 1989 \cite{Atiya:1989tq}. With current luminosity ($\sim 10^{11} K^+$~\cite{CortinaGil:2018fkc}), NA62 should already have recorded at least 100  such events passing the di-muon trigger. Here we propose a di-muon resonance search in $K^+ \to \mu^+ \nu X(\mu^+ \mu^-)$ events with opposite sign (OS) di-muon pairs.  

Since these data have not been released yet by NA62, we estimate the sensitivity of the search from our MC simulation. 
We implement the effective weak interaction of Eq.~\eqref{effectivInt-K}, the electromagnetic interactions of $K^{+}$ decays, and  the new physics couplings from Eqs.~(\ref{eq:lagrangian}) and (\ref{eq:yukawa}) in MadGraph 5 v2 LO \cite{Alwall:2011uj,Alwall:2014hca}. We neglect a subdominant contribution from a contact interaction of $K^+\mu^+\nu\gamma$. 
Both the  background and the signal in $2 \mu^+  \mu^- \nu$ final state are simulated. In Figs \ref{fig:money} (left) and \ref{fig:scalar} (left) we present the results of this analysis in blue curves labeled NA62 $K\to 3\mu \displaystyle{\not}{E}$. Systematic uncertainties on the background will affect less the result compared to the invisible channel because a data-driven background estimate would be possible. For more details about our projection, see supplementary material~\cite{supplement}.

\section{Conclusion}\label{sec:conclusions}

In this Letter we have shown that rare kaon decay searches at NA62 can probe most of the remaining parameter space for which muonic-philic particles resolve the $\sim 3.5\sigma$ $(g-2)_\mu$ anomaly; these are the only viable explanations involving particles below the weak scale. The same parameter space can also accommodate thermal DM production or reduce the $H_0$ tension if the new particle decays
 to DM or neutrinos, respectively.

If this new particle decays invisibly, achieving this sensitivity requires a dedicated single muon trigger to record all $K^+ \to \mu^+$ +invisble events with $m_{\rm miss}^2 > 0.05\text{ GeV}^2$ during Run 3. The ultimate reach in this channel depends crucially on the systematic uncertainties on events with these kinematics; a dedicated experimental study is needed to assess the feasibility of this requirement. 

We note that if the $(g-2)_\mu$ anomaly is confirmed, NA62 can play a key role in deciphering the new physics responsible for the discrepancy. However, even if future
measurements are consistent with the SM, the searches we propose can still constrain  models for which muonic forces mediate dark matter freeze out. Such measurements can also inform future decisions about proposed dedicated experiments including NA64$\mu$\cite{Gninenko:2014pea}, M$^3$\cite{Kahn:2018cqs}, BDX \cite{Battaglieri:2017qen,Battaglieri:2016ggd}, and LDMX \cite{Berlin:2018bsc,Akesson:2018vlm}.

\begin{acknowledgments}
{\it Acknowledgments} :  
We are grateful to Todd Adams, Wolfgang Altmannshofer, Gaia Lanfranchi, Roberta Volpe 
 and Yiming Zhong for helpful discussions and to Babette D$\ddot{\text{o}}$brich, Evgueni Goudzovski for correspondence about NA62. We also thank Evgueni Goudzovski, Roberta Volpe and Yiming Zhong for useful feedbacks on a preliminary version of the manuscript. Fermilab is operated by Fermi Research Alliance, LLC, under Contract No. DE-AC02-07CH11359 with the US Department of Energy.  GMT is supported by DOE Grant DE-SC0012012, NSF Grant PHY-1620074 and by the Maryland Center for Fundamental Physics. KT is supported by his startup fund at Florida State University. (Project id: 084011-550-042584). The authors thank the KITP where this work was initiated and the support of the National Science Foundation under Grant No. NSF PHY-1748958.
\end{acknowledgments}

\bibliography{Kaon_bound}

\clearpage
\newpage
\maketitle
\onecolumngrid

\begin{center}
\textit{\large Supplemental Material}\\
\vspace{0.05in}
{Gordan Krnjaic, Gustavo Marques-Tavares, Diego Redigolo and Kohsaku Tobioka}
\end{center}
\vspace{0.05in}
\twocolumngrid

\section{Decay Calculation}\label{sec:threebody}
 \renewcommand{\theequation}{A.\arabic{equation}}
\setcounter{equation}{0}

The SM width $K\to \mu\nu$ can be written as 
\begin{equation}
\Gamma(K^{+}\to\mu^+\nu)=\frac{m_K \lambda_\mu^2}{2\pi} \left(1-\frac{m_\mu^2}{m_K^2}\right)^2\ .
\end{equation}
where the coupling
\begin{equation}
\lambda_\mu\equiv 2 G_F f_K \, m_\mu V_{us}\simeq8.7\times 10^{-8},
\end{equation}
sets the typical size of the kaon decay widths considered here. Note that $\lambda_\mu$ has to be proportional to the muon mass because a chirality flip is required to make the amplitude non-zero.   
The  kaon width is $\Gamma_{K^+}=5.3\times 10^{-14}\text{ MeV}$, so $\text{BR}_{K\to\mu\nu}\simeq0.63$. Below  we present the calculation for the squared matrix elements of  
\begin{equation}
K^+(P) \to \mu^+(k) \nu_\mu(q) X(\ell)\, ,
\end{equation}
where $X = V$ or $\phi$ is a muonic force carrier considered in this paper and $P, k, q$ and $\ell$ are four vectors. These results are already present in the extensive literature on muonic forces (see for example \cite{Carlson:2012pc}) but we present them here for completeness. 

For either scenario,  the partial width for this process can be written as
\be
\Gamma_{K \to \mu\nu_\mu X} =   \frac{1}{256 \pi^3 m_K^3} \int \sum |  \mathcal{M}_X |^2 d \mone^2 d\mtwo^2\,~,~~~~~~
\ee
where the limits of integration are given by $(\mone^2)_\text{min} = m_X^2$ and $(\mone^2)_\text{max} = (m_K - m_\mu)^2$. For a 
fixed $\mone$ the minimum and maximum of $\mtwo$ are given by
\be
\hspace{-0.1cm}(\mtwo^2)^\text{\small min}_\text{\small max} \! = \! (E_2^* \! + \! E_3^*)^2 \! -  \! \! \biggl( \! \! \sqrt{E_2^{* 2} \!- \! m_{X}^2} \! \pm \! \sqrt{E_3^{* 2}  \! - \! m_\mu^2} \biggr)^{\! 2} \!\!, ~~~~~~~~
 \label{eq:m23limit}  
\ee
where we define
\be
	E_2^* =  \frac{\mone^2 + m_X^2}{2 \mone}~~,~~
	E_3^* =  \frac{m_K^2 - \mone^2 - m_\mu^2}{2 \mone}~.~~
\ee
In Fig.~\ref{fig:signal} we plot for completeness the normalized signal rates for both the vector and the scalar model.

\begin{figure}[t!]
\hspace{-0.2cm}
\includegraphics[width=0.4\textwidth]{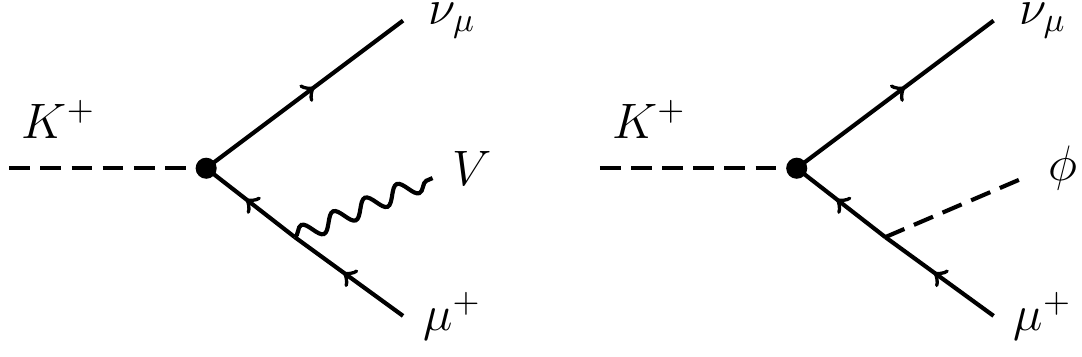}
\caption{Two representative Feynman diagrams that contribute to rare kaon decays involving a light, invisibly decaying vector from Sec.~II (left) and scalar
	from Sec.~III (right). In the vector case there is another diagram where the vector radiates off from the neutrino line. This is not shown but it is included in our result. 
	  }
\label{fig:feyn}
\end{figure}

\subsection{Vector Mediator}
For the vector model introduced in Sec.~II with $X = V$, our process of interest arises from
the Feynman diagram in Fig.~\ref{fig:feyn} and also contains an additional diagram with $V$ emitted from
the $\nu_\mu$.  
The squared matrix element is 
\be
 |\mathcal{M}_V|^2 &=&  g_{V}^2 \lambda_\mu^2 \biggl[ 2 +\frac{(\mone^2+2 m_\mu^2 - 2m_K^2)}{\mtwo^2-m_\mu^2}
   \nonumber \\           
&&\hspace{-1.cm}
	 - \frac{(m_K^2-m_\mu^2)(m_V^2+2m_\mu^2)}{(\mtwo^2-m_\mu^2)^2} 
	 + 2 \frac{(m_K^2-m_\mu^2)^2+m_V^2 m_\mu^2}{\mone^2 (\mtwo^2-m_\mu^2)}
\nonumber \\ 
&&
\hspace{-1cm}
 -  \frac{m_V^2(m_K^2-m_\mu^2)}{\mone^4} 
  + \frac{(\mtwo^2+m_\mu^2-2m_K^2)}{\mone^2}
 \biggr],~~~\label{eq:matrixsqV}
\ee
where $k\, q$ and $l$ are respectively the $\mu, \, \nu$ and $V$ momenta and we define $m_{12} = (\ell + q)^2 $ and $m_{23} = (\ell + k)^2$. Note that the full matrix element vanishes for $m_\mu\to0$ due to chiral symmetry.

\begin{figure}[t!]
\hspace{-0.2cm}
\includegraphics[width=0.46\textwidth]{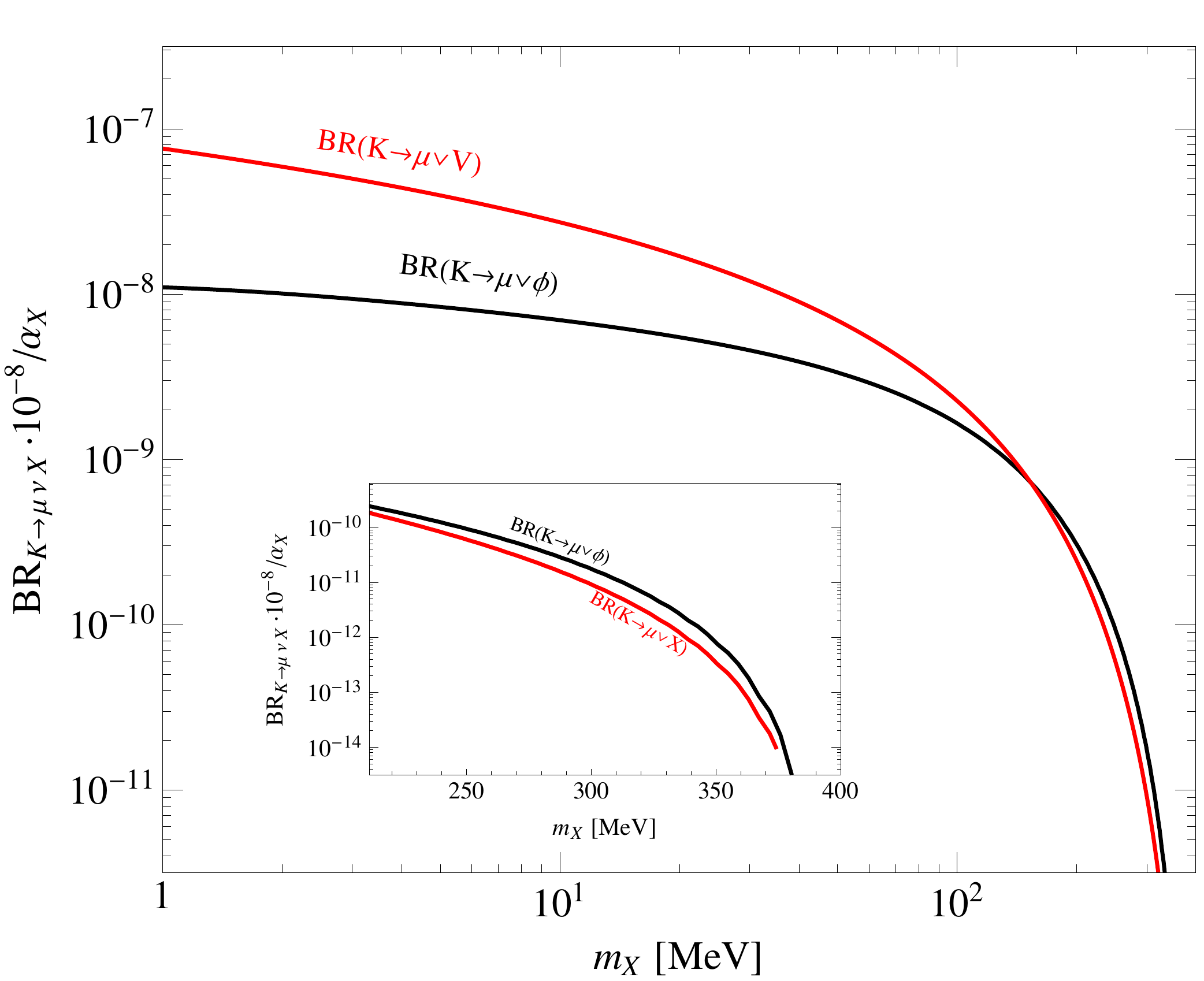}
\caption{
Total branching ratio for $K\to\mu\nu X$ where $X$ is a vector $V$ (red) or a scalar $\phi$ (black) as a function of the the mass of $X$. In the small quadrant we give a zoom of the relevant region for $K\to\mu\nu X(2\mu)$.  
}
\label{fig:signal}
\end{figure}

\subsection{Scalar Mediator}
For the muon-philic scalar introduced in Sec.~III, the squared matrix element  is 
\be
	|\mathcal{M}|^2 &=& \frac{ \lambda_\mu^2  y_\phi^2 }{2m_\mu^2(m_{23}^2-m_\mu^2)^2} \biggl[ m_K^2(\mtwo^2+m_\mu^2)^2 
	\nonumber  \\ 
	&& \hspace{-1cm}
	  -\mtwo^2 \left( (\mtwo^2+m_\mu^2)^2 +\mone^2(\mtwo^2-m_\mu^2)\right) \nonumber  \\ 
	&& \hspace{-1cm}
	   +m_\phi^2 (m_{23}^2-m_\mu^2 m_K^2)\biggr]\ ,
\ee
where $m_{23}$ is defined below Eq.~\eqref{eq:matrixsqV}. The squared matrix element above does not vanish for $m_\mu\to0$ because the scalar yukawa interactions with the muons in Eq.~(4) breaks the chiral symmetry independently of the muon mass. 

\begin{figure*}[t!]
\hspace{-0.2cm}
\includegraphics[width=0.44\textwidth]{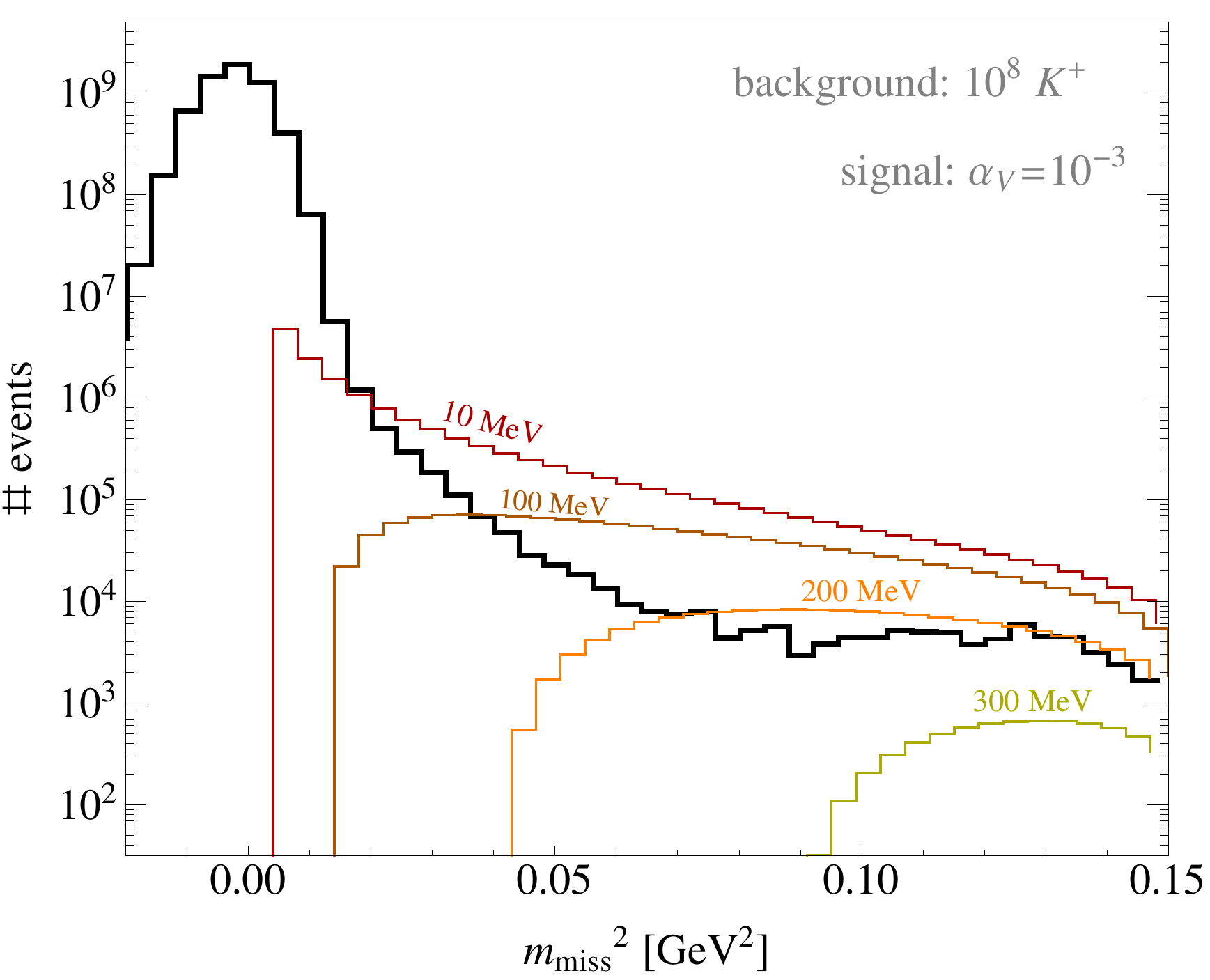}~~
\includegraphics[width=0.44\textwidth]{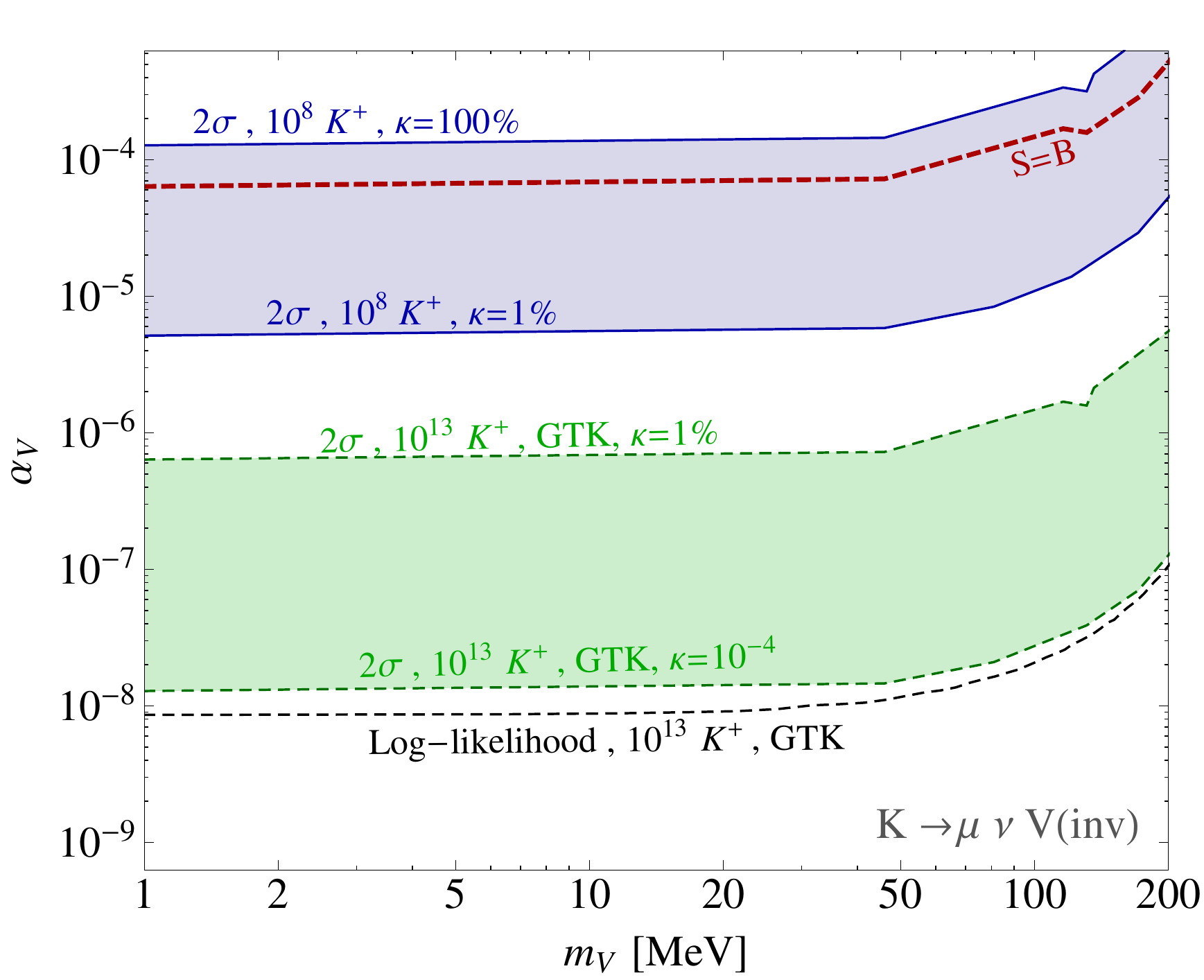}
\caption{{\bf Left:} Missing invariant mass distribution for $K\to \mu \nu V$ decays for different masses of $V$ (in different colors) where $m_{\rm miss}^2$ is the combined invariant mass
of $V$ and $\nu_\mu$ in  Eq.~(10). The missing mass distribution is very similar. In the scalar case very similar distributions are obtained. The black line correspond to the background distribution extracted from \cite{CortinaGil:2017mqf}. The data are binned in squared invariant mass bins of $4\times 10^{-3}\text{ GeV}^2$. {\bf Right:} Sensitivity at $2\sigma$ level of the invisible search for modification of the missing mass tail from $K\to\mu\nu V(V\to\text{invisible})$. The red dashed line shows when the signal is equal to the background extracted from the 2015 data after applying the missing mass cut. The blue band is the present sensitivity based on $10^8$ kaons collected in 2015; the thickness of the band encompasses different assumptions about the magnitude of background systematic uncertainties. The green band shows the future sensitivity based on $10^{13}$ kaons with different systematics. A background suppression at large missing mass is assumed to account for the GTK installation. The dashed black line is based on the likelihood analysis described in Sec. IV, here the background uncertainty is assumed to be dominated by statistics. 
}
\label{fig:signal-bginv}
\end{figure*}

\section{Complete Scalar Model} \label{sec:complete-model}
 \renewcommand{\theequation}{B.\arabic{equation}}
\setcounter{equation}{0}

Before electroweak symmetry breaking, the Yukawa interaction in Eq.~(4) is forbidden by gauge symmetry. The simplest gauge invariant operator that gives rise to this Yukawa interaction is the dimension 5 operator
\begin{equation}
	\phi L H \mu^c ,
\end{equation}
where $L$ is the second generation lepton doublet and $\mu^c$ the muon singlet in 2-component spinor notation.

In this section we present a UV completion of the model which gives rise to this interactions after integrating out heavy fermionic degrees of freedom~(see e.g.~\cite{Chen:2015vqy,Batell:2016ove,Batell:2017kty} for other alternatives). This construction differs from the ones in~\cite{Chen:2017awl} in that the coupling to muons does not arise due to the scalar mixing with the Higgs. In particular the Higgs-scalar mixing is loop suppressed in this model and can be parametrically smaller in a technically natural manner;
thus, as discussed below, many of the scalar bounds presented in \cite{Batell:2016ove} do not apply for an equivalent $\phi-\mu$ coupling. 

The model includes an extra vector-like pair of fermions in which one of these carries the same gauge quantum numbers as $\mu^c$ and 
the other carries compensating quantum numbers to cancel anomalies. 
This extension can generate the required coupling through mixing between this new fermion and the muon. The relevant terms in the Lagrangian of the model are 
\begin{equation}
	\begin{aligned}
	\mathcal{L} & \supset  y_\mu L H \mu^c + M \psi \psi^c + \lambda_1 \phi \psi \psi^c + \lambda_2 \phi \psi \mu^c \\ 
	&+ y_\psi L H \psi^c +h.c. \, ,
	\end{aligned}
	\label{eq:appendix-scalar-model}
\end{equation}
where $(\psi,\psi^c)$ is the new vector like fermion pair. Note that we chose to not include a mass mixing term $\mu^c\psi$ which is allowed by all the symmetries, since this term can be removed by an appropriate field redefinition. Also there are additional terms such a tadpole $\lambda_1 \phi M^3$ or a cubic term $\lambda_1 \phi |\phi|^2 M$ which shifts mass of $\phi$ or mix Yukawa couplings. However these can be avoided by small values of $\lambda_1$ which is allowed by spurion analysis.  

Assuming $M>v$, we can integrate out the new fields before electroweak symmetry breaking. This generates the following new terms
\begin{equation}
	-\lambda_2 y_\psi \frac{\phi}{M} L H \mu^c + \frac{\lambda_1 \alpha}{6 \pi M} \phi F^{\mu \nu}F_{\mu\nu},
	\label{eq:integrate-out-fermion}
\end{equation}
where $F$ is the photon field strength. After electroweak symmetry breaking the first term in the above interaction generates the coupling in Eq.~(4), with  
\begin{equation} 
	y_\phi = -\frac{\lambda_2 y_\psi v}{\sqrt{2} M} \, .
\end{equation} 
The second term in Eq.~(\ref{eq:integrate-out-fermion}) contributes to the scalar decay to photons. Depending on the choice of parameters the contribution from this term can be larger than the IR contribution from the muon loop and  the partial width to photons in Eq.~(6) must be corrected. This shows that different choice of UV parameters can lead to either prompt or displaced decays to photons, which highlights the complementarity of performing both an invisible search and a diphoton resonance search.

The couplings in Eq.~(\ref{eq:appendix-scalar-model}) induce a $\phi$-Higgs mixing at loop level. We can estimate the size of the mixing from the contributions involving $\lambda_1$ and $\lambda_2$ to be
\begin{equation}
	\frac{\lambda_1 y_\psi^2}{16 \pi^2} M v \phi h \, ,  \ \ \ \frac{\lambda_2 y_\psi^2}{16 \pi^2} M v \phi h \, .
\end{equation}
This induces mixing angles much smaller than the $\sim 10^{-3}$ bound discussed in~\cite{Krnjaic:2015mbs,Batell:2016ove} for all of the parameter space we are interested in. One can also easily show that the decay to electrons induced by the mixing with the Higgs is small compared to the diphoton decay from the $\phi F^2$ coupling.

\section{NA62 Analysis}\label{app:analysis}
 \renewcommand{\theequation}{C.\arabic{equation}}
\setcounter{equation}{0}

Here we provide additional details about the procedure with which NA62 projections are computed this paper. In particular, we present the background distributions for both the visible and invisible analyses and comment on how different assumptions regarding systematic errors affect these projections.
Maximizing signal sensitivity is challenging for two main experimental reasons: 

\begin{enumerate}
\item{\bf Single muon trigger bandwidth:}  This issue is related to the large number of single muon events from SM $K\to \mu\nu$ decays. Thus, the current single muon trigger at NA62 is rescaled by $1/400$ \cite{CortinaGil:2018fkc}, so only one single muon event out of  400 is recorded, which reduces the sensitivity of our search. 
This limitation can be overcome with a dedicated single muon trigger with a lower cut on the missing mass at trigger level (or equivalently an upper cut on the muon momentum). In the 2015 data sample, despite over $2.4 \times 10^7$ events passing the single muon trigger, only $5.6\times 10^3$ have $m_{\text{miss}}^2>2.3\times 10^{-2}\text{ GeV}^2$. Thus, with a dedicated trigger, it would be possible to record \emph{all} events with $m_{\text{miss}}^2>2.3\times 10^{-2}\text{ GeV}^2$ and keep the $1/400$ trigger rescaling for those with lower $m_{\text{miss}}$. Our search strategy exploits this possibility  and utilizes the full NA62 luminosity $N_{K^+} \approx 10^{13}$ in the decay region, which we assume for our projections. As a final remark, notice that our signal region has kinematical overlap with the region 2 of the $K^{+}\to\pi^{+}\nu\bar{\nu}$ search \cite{CortinaGil:2018fkc}. A modification of the single muon trigger prescaling could then possibly affect the background yield from muons faking pions in the R2 region. 
\item{\bf Background systematics for large $m^2_{\rm miss}$:}  These systematics are difficult to estimate from the 2015 data in which there is disagreement between data and Monte Carlo (MC)  at large $m_{\text{miss}}$.  A careful experimental effort is required to assess these uncertainties. Since our goal is to show how much the sensitivity of NA62 could potentially be improved, we presents results with only statistical errors; these can only be achieved once systematic uncertainties become subdominant for the full NA62 luminosity: ${\sigma_{
\rm sys}}/{B}< B^{-1/2} \sim  10^{-4}$. 
In Figs.~1 and 2 we presented future sensitivities assuming systematics are negligible, but note that exploring 
new parameter space in this plane only requires systematic uncertainties to be below $1 \%$.  
\end{enumerate}

\subsection{Invisible analysis}
\label{invisible-appendix}
 
 In Fig.~\ref{fig:signal-bginv} left we compare the $m^2_{\rm miss}$ distribution for $K\to\mu\nu X$  signal events for different $X$ masses using the background shape extracted from NA62 public data \cite{CortinaGil:2017mqf}. The signal here is shown for $X=V$ but the scalar case is qualitatively similar. Note that the signal reduction at small $m_{\text{miss}}^2$ is  $m_X$ dependent, so an optimal $m_{\text{miss}}$ can be chosen for different values to maximize sensitivity. As discussed in Sec.~IV A, the background at large missing mass does not appear to scale as one might expect if it were dominated by the QED radiative tail from $K\to\mu\nu(\gamma)$ decays. The reason is that other backgrounds including the halo muon background and $K\to 3\pi$   become dominant in  this regime. We believe that these backgrounds will be further suppressed in future data releases for which timing and momentum of the kaon will be measured upstream with the silicon pixel detector (GTK), which has already been used for the 2017 run. To roughly account for this improvement, we rescale the background above $m_{\text{miss}}^2>0.023\text{ GeV}^2$ by an additional factor of two. 

In Fig.~\ref{fig:signal-bginv} right we show estimated $2\sigma$ sensitivities for the vector case computed in a cut-and-count experiment; similar results are also found for the scalar case. This simpler analysis is performed here and compared to the likelihood analysis presented in the main text in order to quantitatively show the effects of systematic uncertainties on the background. 

The $2\sigma$ sensitivity of an $m^2_{\text{miss}}$ search in single muon events is computed by evaluating $S /\sqrt{B+\kappa^2 B^2}=2$, where the $S$ is the number signal events, $B$ the number of background events and $\kappa=\sigma_{
\rm sys}/B$ is the systematic uncertainty on the background. The signal yield is 
\begin{align}
S= \frac{ N_{K^+}  \, \mathcal{A} }{\Gamma_{K^+}} \int_{m_{\text{cut}}^2}^{m_{\text{max}}^2} dm_{\rm miss}^2 \frac{d \, \Gamma_{K^+\to \mu^+\nu X}}{ dm_{\rm \tiny miss}^2} \, , \label{eq:signalinv}
\end{align}
where $\mathcal{A}\simeq0.35$ is the the detector acceptance. $m_{\text{cut}}$ is the lower cut on the missing mass, which is optimized for each value of $m_X$ to maximize signal sensitivity, but always satisfies $m^2_{\rm cut } > 0.05 \text{ GeV}^2$; $m_{\text{max}}^2=(m_K-m_\mu)^2=0.15\text{ GeV}^2$ is the maximum kinematically allowed missing mass.\footnote{Note that $m^2_{\rm cut }=0.05 \text{ GeV}^2$ is the \emph{minimal} missing mass cut in our cut and count analysis. This should not be confused with $m^2_{\rm miss}=0.023 \text{ GeV}^2$ which is the value of the invariant mass below which adding bins to the log-likelihood ratio in Eq.~(11) does not effectively improve signal sensitivity. Of course the physics behind these two quantities is very similar and related to the background shape peaking at $m_{\text{miss}}=0$.}  

From Fig.~\ref{fig:signal-bginv} it is clear that the future NA62 sensitivity depends greatly on background systematics at large missing mass. For the present/future luminosity, the blue/green lines at the bottom of these bands correspond to systematic uncertainty $\kappa_{\text{min}}$ for which the statistical uncertainty becomes dominant. This can be estimated as a function of the luminosity and the number of background events for a given missing mass cut
\begin{equation}
\kappa_{\text{min}}\simeq \frac{1}{\sqrt{B} }\simeq (4\times 10^{-2}, 1.8\times 10^{-4}),
\end{equation}
where the first number assumes $10^{8}$ kaons and 652 background events after a missing mass cut of $m_{\text{miss}}^2>0.05\text{ GeV}^2$ and the second number assumes $10^{13}$ kaons and $3.2\times 10^7$ background events (accounting for the expected background suppression). For comparison we also show in Fig.~\ref{fig:signal-bginv} the most aggressive reach derived from our likelihood analysis. As expected, the log-likelihood improves the reach for low mass resonances where the signal spreads widely in the large background region (see Fig.~\ref{fig:signal-bginv} left) and a simple cut-and-count analysis poorly distinguishes 
the signal from background.  
\begin{figure}[t!]
\hspace{-0.2cm}
\includegraphics[width=0.46\textwidth]{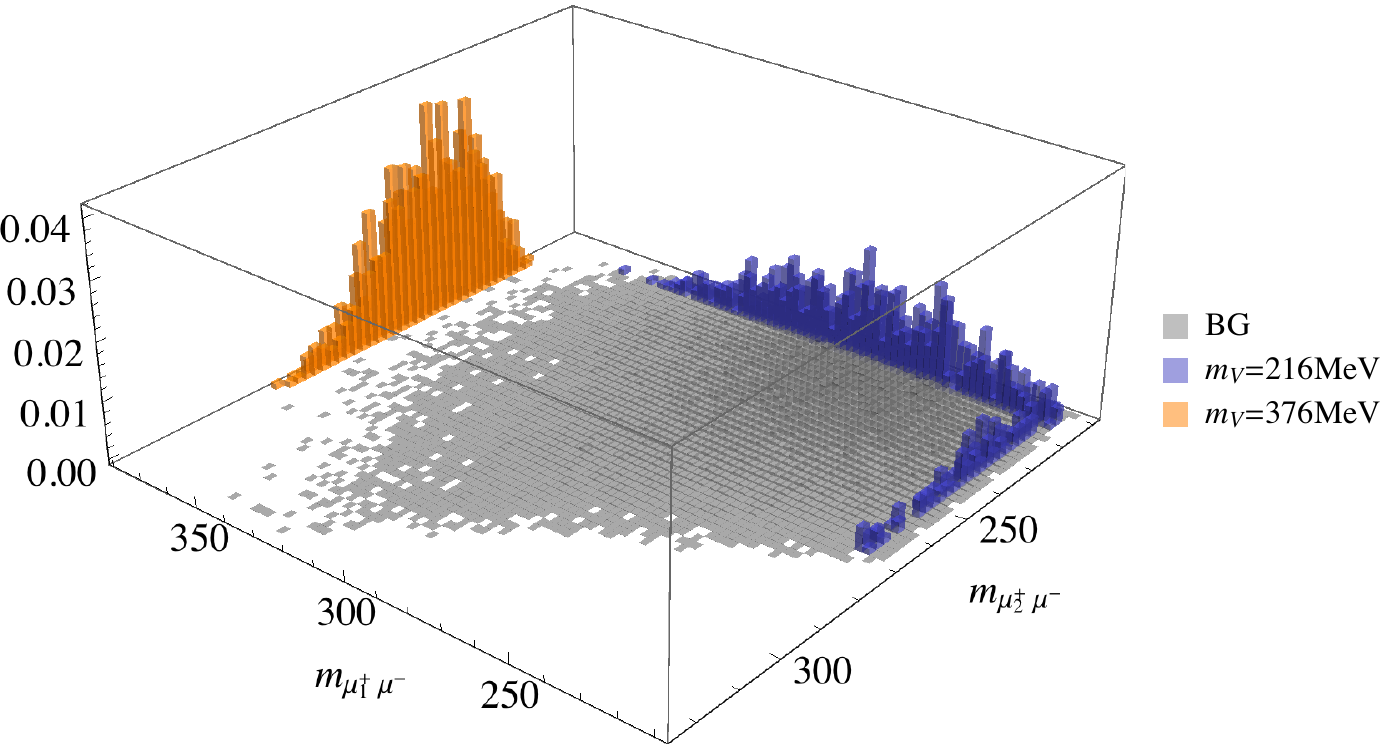}
\caption{   
Normalized 2D distributions of $m_{\mu_1^+\mu^-}$ and $m_{\mu_2^+\mu^-}$ for signal of two benchmark points ($m_V=216, 376$~MeV) and the SM background. The muon momenta are evaluated in the $K^+$ rest frame, and $\mu_1^+$ corresponds to the leading muon and $\mu_2^+$ is the other. As expected, $\mu_1^+$ leads to a peak of $V(\phi)$ for the higher mass of the muonic force, while $\mu_2^+$ does the same for the lower mass. 
}
\label{fig:signal-bg2muon}
\end{figure}

\subsection{Di-muon analysis} \label{dimuon-appendix}

In this section we describe the proposed opposite-sign di-muon resonance analysis in $K\to 3 \mu \nu$ events, which defines the blue projections in Fig. 1 (left) and 2 (left) labeled NA62 $K \to 3 \mu + \displaystyle{\not}{E}$.
We assume that the irreducible SM background for our search arises from $K^+$ decays to three muons through an off-shell gauge boson and neglect other possible backgrounds from non-detection of photons, $\pi^\pm$ misidentification and decay which are expected to be in the same order or subdominant. Moreover, for simplicity, we assume acceptance for both signal and background to be 5\%. This number is roughly 1/6 of the acceptance reported in \cite{CortinaGil:2017mqf} for the single muon trigger and should roughly account for the extra cost of requiring three muons to pass trigger and identification criteria. 

The other challenge of this search is the ambiguity in choosing the opposite-sign di-muon pair to reconstruct the $X$ invariant mass. To resolve this problem we choose the opposite-sign di-muon pair that gives an invariant mass closer to each test mass for the signal. 
Typically it is the leading muon above $m_X\simeq300$~MeV, and the second leading one below $m_X\simeq260$~MeV as seen in Fig.~\ref{fig:signal-bg2muon}.  
After this choice is made, we select the signal and the background within a narrow invariant mass bin around each test mass $[m_{X}-2\delta m_{\mu^+\mu^-}, m_{X}+2\delta m_{\mu^+\mu^-}]$. The invariant mass bin size can be determined as a function of the smearing of the muon momentum in the NA62 detector
\begin{equation}
\frac{\delta m_{\mu^+\mu^-}}{m_{X}}=\frac{1}{2}\left(\frac{\delta p_{\mu_{+}}}{p_{\mu_+}} \oplus\frac{\delta p_{\mu_{-}}}{p_{\mu_-}}\right)
\end{equation}
where $\delta p_{\mu}$ is the muon momentum resolution of the NA62 detector, which satisfies \cite{CortinaGil:2017mqf}
 \begin{align}
\frac{\delta p_{\mu}}{p_{\mu}}=0.3\%\oplus \left( 0.005 \, \frac{p_{\mu}}{\rm GeV} \right) ,
\end{align}  
where $\oplus$ indicates a sum in quadrature. The muon momentum is fixed to be $p_{\mu_{\pm}}=20~\GeV (\sim p_{K^+}/4)$. 
The future sensitivities of this search at 2$\sigma$ in Fig.~1 (left) and Fig.~2 (left) assume $10^{13}$ kaons and uncertainties dominated by statistics. Systematic uncertainties on the background can be under control because the data-driven background estimate (side-band) is made possible by the peaked nature of the signal.


\end{document}